\documentclass[sn-apa, iicol]{sn-jnl}

\jyear{2022}%

\theoremstyle{thmstyleone}%
\theoremstyle{thmstyletwo}%

\theoremstyle{thmstylethree}%

\begin{document}

\title[Article Title]{Sequence-aware item recommendations for multiply repeated user-item interactions}


\author[1,3]{\fnm{Juan Pablo} \sur{Equihua}}\email{je18890@essex.ac.uk}

\author[2]{\fnm{Maged} \sur{Ali}}\email{maaali@essex.ac.uk}

\author[3,1]{\fnm{Henrik} \sur{Nordmark}}\email{henrikn@profusion.com}


\author*[1,4]{\fnm{Berthold} \sur{Lausen}}\email{berthold.lausen@fau.de}

\affil[1]{\orgdiv{Department of Mathematical Sciences}, \orgname{University of Essex}, \orgaddress{ \city{Colchester}, \country{United Kingdom}}}

\affil[2]{\orgdiv{Essex Business School}, \orgname{University of Essex}, \orgaddress{ \city{Colchester}, \country{United Kingdom}}}

\affil[3]{\orgdiv{Innovation}, \orgname{Profusion Media Ltd.}, \orgaddress{\street{Paul Street}, \city{London}, \country{United Kingdom}}}

\affil[4]{\orgdiv{Institute of Medical Informatics, Biometry and Epidemiology, School of Medicine}, \orgname{Friedrich-Alexander University Erlangen-Nuremberg},
    \orgaddress{\street{Waldstr. 6}, \city{Erlangen}, \postcode{91054},  \country{Germany}}}



\abstract{Recommender systems are one of the most successful applications of machine learning and data science. They are successful in a wide variety of application domains, including e-commerce, media streaming content, email marketing, and virtually every industry where personalisation facilitates better user experience or boosts sales and customer engagement. The main goal of these systems is to analyse past user behaviour to predict which items are of most interest to users. They are typically built with the use of matrix-completion techniques such as collaborative filtering or matrix factorisation. However, although these approaches have achieved tremendous success in numerous real-world applications, their effectiveness is still limited when users might interact multiple times with the same items, or when user preferences change over time. 

\vspace{0.5cm}
We were inspired by the approach that Natural Language Processing techniques take to compress, process, and analyse sequences of text. We designed a recommender system that induces the temporal dimension in the task of item recommendation and considers sequences of item interactions for each user in order to make recommendations. This method is empirically shown to give highly accurate predictions of user-items interactions for all users in a retail environment, without explicit feedback, besides increasing total sales by  5\% and individual customer expenditure by over 50\% in an A/B live test.}



\keywords{Recommender systems, Natural language processing, Deep learning, Customer Experience.}



\maketitle

\section*{Introduction}\label{sec1}

Recommender systems are perhaps one of the most successful applications in Machine Learning (ML) and Artificial intelligence (AI) in the last two decades. These systems have been designed, tested, and implemented successfully in a wide range of application domains to expose users to a large collection of relevant items, and are particularly helpful when the catalogue of potentially recommendable items is large enough to make it impossible to do recommendations manually for humans. Recommender systems in general are recognised as an efficient approach to provide users with personalised content. For example, the streaming service Netflix displays user-level predicted movie ratings to its customers to help them in deciding which content is more suitable and interesting to watch. The global online retailer Amazon provides predicts item ratings to users based on the previous purchase history of similar customers. As the final goal of these tools is to suggest which items are more suitable to individual users or which items will be liked by customers, these systems are typically categorised as recommender systems. 

Recommender systems have their roots in the field of information retrieval; identifying which online written content might be most relevant with respect to a given user query, and then sorting the retrieved list of top relevant documents based on easy user consumption. However, although this approach was widely used in the ‘90s by different companies for recommending specific items, it was quickly replaced with more advanced techniques. These include content-based Collaborative Filtering (CF) \cite{cf1} and Matrix Factorization (MF) \cite{Koren_2009} recommender systems, which aim to learn a relationship between user preferences and items by using historical information of user’s actions and purchases in a matrix-completion framework, such as Singular Value Decomposition (SVD) \cite{ZHOU_2015}, where the goal is to predict future user preferences in a user-item rating matrix. These well-established techniques also have their strengths and weaknesses, and many researchers, companies, and AI practitioners have chosen to combine techniques in different ways to provide better recommendations for users and increase either overall revenue, customer engagement, or model performance. This led into the development of Hybrid Recommender systems \cite{Burke_2002}, where techniques such as weighted recommenders \cite{claypool99}, mixed recommenders \cite{SMYTH200053},  and feature combined recommenders \cite{Basu1998} are perhaps the most popular methods used in the industry. Although these techniques usually can deal with some of the issues in recommender systems, such as the cold-start problem of adding new items or users into the system \cite{Jazayeriy_2018}, they still have several performance issues in several application domains \cite{MARCHAND2020328}.

Across industry, recommender systems are powerful tools to enhance user experience via personalisation and increase sales and overall revenue by identifying which items are most likely to be relevant for users \cite{ RS_Textbook,Gunawardana_2009, Zhao_2014}. Mirroring its use in other applications such as computer vision and natural language processing (NLP), deep learning is capable of great achievements in the field of recommender systems, where uncovering non-linear complex relationships between user-items interactions with the use of deep neural networks can easily outperform longer-standing techniques, and these models are capable of learning complex user-item relationships from the usually-abundant data itself \cite{Zhang_2019}. Implementation of deep learning for recommender systems have proven to significantly outperform other techniques without requiring major efforts at the deployment stage. Further detail of these kind of approaches is included in the third section.

Despite techniques such as CF and MF empowered with the use of deep learning have achieved tremendous success in real-world applications at being able to capture nonlinear user-item relationships, these still have their caveats and tend do not perform well when users’ preferences change over time, or when users interact several times with only a few items within the total catalogue, which is commonly the case for many specialised retailers and small and medium enterprises (SMEs) settings. Furthermore, as CF and MF are matrix completion approaches, these commonly make the assumption that the exact user preference of items, a.k.a user rating, is known for all user-item interactions in the data. There has been research to propose synthetic ratings for user-items pairs \cite{Xiaoyuan_2009, Sidana_2017}, such as by creating functions which only depend on how many times users have purchased or interacted different items. The construction of these synthetic ratings is often completely arbitrary and can lead to unrealistic assumptions and inaccurate predictions when the systems are tested in real-world scenarios . 

Over the last few years, there has been a vast amount of research around sequence-aware recommender systems \cite{quadrana2018, Chao_yuan_2017, Pereira_2021, Shani_2005, Zhang_2019} particularly in the context of implicit feedback, which is when the exact user rating of items is unknown for all or most user-item interactions in the data. Taking inspiration from NLP, sequence-aware recommender systems are inherently different from the traditional matrix-completion approaches. Sequential methods process customer transactions as sequential information by considering each item in the catalogue of products as a single word in a dictionary or token \cite{quadrana2018}, and taking all transactions made by each customer to build a sequence of tokens, similarly to processing sequences of words. Then, the new sequences generated for each user can be used to embed customers into a common representation space and make predictions of the most likely token or item to appear next in the sequence, and item recommendations can be as simple as selecting the most next likely item to be part of the sequence of purchased items. 

The rest of the paper is organised as follows: the second section reviews the current literature of recommender systems and how these get combined with deep learning and neural networks and sequential models. Third section outlines the methodology of our sequence-aware recommender system and how transactions can be compressed and expressed as sequences of tokens. Fourth section presents the experimental results carried with two large retailers in the United Kingdom and the \emph{MovieLens} dataset commonly used for recommender systems benchmark. Finally, Fifth section outlines the main conclusions of this work and future lines of research.

\section*{Existing work}\label{section2}

\subsection*{Notation}\label{notation}

In the typical recommendation setting, there is a set of users $U = \{u_{1},u_{2}, ..., u_{ \mid U \mid } \}$  with size $\mid U \mid $, and a set of items $I = \{i_{1},i_{2}, ..., i_{ \mid I \mid }  \}$ with size $\mid I \mid $, although most of the time due to business constraints, it is useful to focus on ranking only a subset of potentially recommendable items $I’ \subset I$, due to the fact that items might not be available at all times or are discontinued from stock. Generally speaking the goal of a recommender system is to produce a list of \emph{relevant} items $L_{u_{j}} \subset I$ for each user $u_{j} \in U$. This is typically achieved by learning a mapping function $f(X_{U}(u_{j}), X_{I}(i_{k})): X_{U} \times X_{I} \rightarrow \mathbb{R}$ that can assign a prediction of how \emph{relevant} the item $i_{k}$ is for user $u_{j}$, where $X_{U} \text{ and } X_{I}$ represent the feature space of the total information available for users and items respectively, such as  user's ratings, user and item characteristics, and hand-crafted features from data. 

Once predictions are computed, the recommendation list $L_{u_{j}}$ can be obtained by sorting items from highest to lowest \emph{relevancy} for each user $u_{j}$. The following sections explore different methods used in research and industry applications to learn this function $f$ from implicit feedback settings. 


\subsection*{Implicit Feedback Recommender Systems.}\label{section2_1}

In the early 90’s, recommender systems relied heavily on the use of explicit feedback of recommendations collected from user’s ratings and reviews for each item \cite{cf1, Resnick_94, claypool99}, which typically are a clear indicative of the likeness or preference to a product of its characteristics, and an indicator of customer loyalty \cite{Ravula_2022}. However, with the large scaling and overwhelming use of recommender systems by companies, obtaining reliable explicit feedback is getting more difficult over time. Furthermore, for some domains, it is inherently difficult or impossible to obtain explicit feedback from users. For example, in the retail environment users typically include a large number of items in their shopping basket for a single visit to the store. Asking a user to provide feedback for each of the items after the purchase was made would not be practical for the user or beneficial in terms of user experience. In this case, we must rely solely on implicit feedback. This may be obtained as a pseudo-rating or estimated with a function typically defined arbitrarily during the data pre-processing phase and computed from different data user’s signals, such as number of purchases for an item or different types of events carried during an online session. 

Several implicit feedback-based recommender systems have been proposed due to the scarcity of reliable explicit ratings provided by users. For instance,  \cite{Valdez_2019} explore different alternatives to recommend electronic books using implicit feedback obtained from the logging information of users in an e-commerce platform, such as duration of the session, number of clicks, users reading time, number of comments.  \cite{hu_2008} proposed obtaining information about the positive or negative preference of users associated via an association with varying confidence intervals to recommend television shows at large scale. \cite{LEE_2008} performed item recommendation on an e-commerce platform with the use of a pseudo-rating and introducing temporal information by including the time that users interact with items and the time elapsed since the item was initially introduced in the platform, affecting the recommender accuracy by promoting brand new items to users.

\subsection*{Matrix-Completion Recommender Systems}\label{section2_2}

\subsubsection*{Collaborative Filtering}\label{section2_2_1}
\begin{figure}[!htbp]
\centering{}
\includegraphics[scale=0.50]{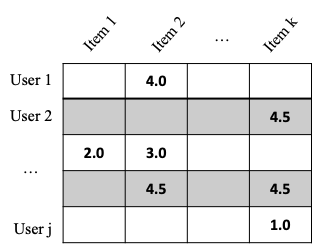}
\caption{User-item matrix of ratings representation for a set of users $u \in U$ and a set of items $i \in I$}
\label{fig:user_item_matrix}
\end{figure}

Collaborative filtering was proposed in the late 90’s as a convenient alternative to content-based and feature-based recommender systems. These had mostly focused on extracting human-engineered characteristics of users and/or items data, aiming to find similarities between purchases to predict future users’ relevant items. Instead, collaborative filtering methods such as \emph{GroupLens} \cite{Resnick_94} aim to estimate how much a user will like a specific item based on how much a set of users liked similar items previously. In Collaborative filtering, ratings are estimated from the user-item matrix shown in Figure \ref{fig:user_item_matrix}, where the known previous user ratings are denoted as $r_{u,i}$ for user $u$ and item $i$, and future unknown ratings $\hat{r_{u,i}}$ are estimated as 

\[
\hat{r}_{u,i} = \frac{\sum_{j \in I} {r_{u,i} \cdot w_{i,j}}}{\sum_{j \in I} w_{i,j}},
\]
\medskip
where $w_{i,j}$ is the similarity between the item $i$ and an item $j \in I$, which can be obtained via any similarity function such as cosine similarity, Jaccard similarity, KL divergence, among others.
 
In practice, estimated ratings $\hat{r}_{u,i}$ are obtained only from a subset of items $J \subset I$ usually called \emph{neighbourhood of item i} rather than from the whole set of items, this to improve scalability and reduce variability at using only the most similar items. 

Several successful applications in industry have used collaborative filtering to improve sales and revenue. For example, \cite{Linden_2003} deployed an \emph{item-item collaborative filtering} recommender in the large e-commerce \emph{Amazon.com} to personalise the content that is displayed for each of the millions of users visiting the website for shopping on a daily basis regardless of the number of ratings or purchases made by previous customers. Additionally, collaborative filtering methods can be easily combined with a large number of machine learning techniques such as clustering \cite{Ungar1998}, latent semantic analysis \cite{Hofmann_2004}, or Markov decision processes \cite{Shani_2005} to improve rating estimation performance and scalability. 

Although CF techniques are relatively easy to implement in production environments and have proven to improve revenue and customer satisfaction in several applications for different industries, these methods have several limitations. Firstly, the fully known user ratings should explicitly and reliably express users’ preferences for items, which is usually difficult to achieve in implicit feedback settings, as typically the rating is only obtained from users’ signals. Secondly, as future users’ ratings are directly estimated from items similarities $w_{i,j}$ highly sparse datasets where only few items are rated by individuals induce extra variability in estimated ratings, leading into overall lower performance and miss leading recommendations.

\subsubsection*{Matrix Factorisation}\label{section2_2_2}

To overcome the data sparsity issue in recommender systems, different approaches implement dimensionality reduction techniques such as \emph{Singular Value Decomposition} (SVD) and \emph{Principal Component Analysis} (PCA), to compress the highly sparse user-item interactions matrix into a low-dimensional dense representation of users and items. 

Matrix Factorisation \cite{Koren_2009} characterises both users and items by latent factors of dimension $k$ directly inferred from the user-items interactions matrix, in such a way that unknown ratings can be easily estimated by the inner product of users and items latent factors. Mathematically, each item $i \in I$ is associated with a hidden latent vector $q_{i} \in R^{k}$, and each user $u \in U$ to a vector $p_{u} \in R^{k}$, where for each user and item, the corresponding vectors $p_{u}$ and $q_{i}$ measure to what extend the $u$ and $i$ associate to the corresponding latent factors positively or negatively, and typically $k<<min(\mid U \mid, \mid I \mid )$. Thus, the estimated user rating for an item, can be estimated as the \emph{‘similarity’} between user and item latent factors, i.e.,

\[
\hat{r}(u,i) = q_{i}^T \cdot p_{u}
\]
\medskip

At training phase, latent factors can be learnt through minimising the error between observed and predicted ratings for which the rating is fully known, typically by using the mean squared error ($MSE = \frac{1}{\mid U \mid \times \mid I \mid}\sum_{j=1}^{\mid U \mid}\sum_{k=1}^{\mid I \mid}(r_{u_{j},i_{k}}-\hat{r}_{u_{j},i_{k}})^2$) as loss function. Then, recommendations can be made by selecting the predictive ratings for which the inner product of latent factors is the largest. Several pieces of research have extended this concept: \cite{bell_2007} combine a neighbourhood-based collaborative filtering with SVD MF at a higher level to improve estimation performance on large datasets without performing any imputation for missing ratings and avoid parameter shrinkage at training. \cite{ Paterek_2007} proposed a weighted SVD by including additional biases to SVD and additional post-processing via kernel ridge regression for each item.\cite{ Salakhutdinov_2007} introduced probabilistic matrix factorization (PMF) which includes adaptive priors in model parameters to outperform SVD while maintaining model scalability for large datasets even for users with few item interactions.

\subsection*{Deep learning-based recommender systems} \label{section2_3}
As mentioned in the first section, recommender systems and neural networks such as feed-forward neural networks (FNN), convolutional neural networks (CNN), and recurrent neural networks (RNN) can be combined in order to find nonlinear and non-trivial user-items interactions and provide better recommendations to users \cite{He_2017, Covington_2016, Hariri_2012, Zhang_2019}.  These types of architectures typically built from multiple neural building blocks can be defined into single differentiable function, and trained end-to-end with classification or ranking losses to foster recommendation performance and item lists sorting, besides having the ability to incorporate data from multiple shapes like users reviews, tweets, item images, or sound in their input data, aiming to resemble the behaviour and benefits of hybrid-recommender systems while avoiding expensive human-based feature engineering.   

Thanks to the easy accessibility to deep learning frameworks such as \emph{Tensorflow} and \emph{Pytorch}, and the increasing computational power available in modern computers, deep learning-based recommender systems have been applied in several research and commercial applications for different industries over the last decade. For example, \cite{He_2017}, replaced the inner product in collaborative filtering with a multi-layer perceptron (MLP) that can learn an arbitrary function from data and find non-linear relationships in the user-item matrix and outperform several Collaborative filtering methods. \cite{Covington_2016}, used two neural networks combined in a candidate-ranking classification framework to produce highly scalable recommendations for users of the large video streaming platform \emph{YouTube}, and reduce the ranking space from several millions of items to just a few thousand. In this work, authors use information about different types of users’ actions combined with item embeddings to produce a list of relevant items with high precision. \cite{Volkovs_2017} proposed a method called \emph{DropoutNet} that combines matrix factorisation and neural networks to address the cold-start problem under the assumption that not having information of users is similar to handling performance missing data efficiently.

\subsubsection*{Sequence-Aware Recommender Systems} \label{section2_4}

Modern recommender systems deployed in production environments rely significantly on the use of matrix-completion techniques combined with users and item characteristics to detect long-term user preferences. However, most of the time these techniques do not perform well in settings where user preferences can change significantly over time, or when users interact repeatedly with the same set of items. Sequence-aware recommender systems typically consider user-item interactions in sequential frameworks to detect drifts in user preferences over short periods of time, and to identify short-term popularity trends quickly and efficiently \cite{quadrana2018}. Typically, the input for these systems is an ordered and timestamped list of past user actions and the output is an ordered list of items most likely to be relevant for the user, just as in the traditional item recommendation setup explained previously.

 \cite{quadrana2018} categorise the sequence-aware recommendation setting into four different categories: \emph{Context adaptation}, when it is important to understand the context of the user, such as geographic location, the current weather, or the time of day to make relevant recommendations. \emph{Trend detection}, where it is critical to have information about community and individual trends of popular items. \emph{Repeated recommendations}, which is when users might interact repeatedly with each item in a single or multiple sessions. \emph{Order constrained}, where the actual order on which user’s actions were made reveal the inherent most likely action to be taken next by the user. 

Sequence-aware recommenders have been widely researched over the last years for multiple application domains. For instance, \cite{Shani_2005} proposed a recommender system based on a Markov Decision Process (MDP) in a reinforcement learning framework to improve revenue of recommended items in an online bookstore. \cite{Baeza_2015} studied how to improve mobile application usage to provide personalised user experience via prediction of which application is likely to be used in the near future by the user, with the use of popular recommendations and session-based feature engineering authors can outperform different prediction methods like Naïve Bayes and Support Vector Machines (SVMs), besides approaching the cold-start problem of having new apps constantly. \cite{Hariri_2012} presented a context-aware recommender system for music recommendations by analysing sequences of previous songs listened by the user within the current session and a database of human-compiled playlist mapped into sequences of topics, achieving better recommendation performance than collaborative or content-based filtering methods.  \cite{Wang_2013} proposed a repeated-interaction recommender system for the e-commerce industry which combines the proportional hazard assumption from survival analysis to model the joint probability of user interacting with items over a period of time.  

\section*{Methodology} \label{section3}
This section describes the mathematical formulation of modelling user-item interactions over time and the architecture used to estimate how likely users are to interact with a specific set of items $I’ \in I$ over a defined horizon. Sequential modelling techniques are such as Markov Decision Processes (MDP), Latent Dirichlet Allocation (LDA), and Recurrent Neural Networks (RNNs) are useful approaches for tasks where the data to be analysed has a temporal dependency and/or a inherently sequential form, like in natural language process and understanding tasks such as sentiment analysis, text classification, and next token prediction. Although the problem of performing item recommendations for users has typically not been seen as a sequential task, there are several sequential techniques that can be implemented to achieve better recommendation performance. In this work, we aim to process user transactions in a sequential framework using inspiration and data pre-processing techniques from the field of NLP, such as tokenisation and sequential embeddings, in a recurrent neural network to make probability estimations for each user interacting with specific items in the dataset.

Following notation introduced in the second section, we can describe the problem of sequential recommendation as follows, let $U = \{u_{1}, ..., u_{ \mid U \mid } \}$ be the set of users, and $I = \{i_{1}, ..., i_{ \mid I \mid } \}$ the set of items. In contrast to matrix-completion methods presented previously where the goal is estimating the overall user estimated rating $\hat{r}(u_{j},i_{k})$ for each user $u_{j} \in U$ and item $i_{k} \in I$, our goal is finding the probability that the user $u_{j}$ interacts with the item $i_{k}$ in a defined horizon, represented as $P_{u_{j}}(i_{k}) = P(i_{k} \mid Seq(u_{j}))$, where $Seq(u_{j})$ is the ordered sequence of items previously bought by the user $u_{j}$, and its detailed construction is presented in the next section. 

\begin{figure*}[!htbp]
\centering{}
\includegraphics[scale=0.60]{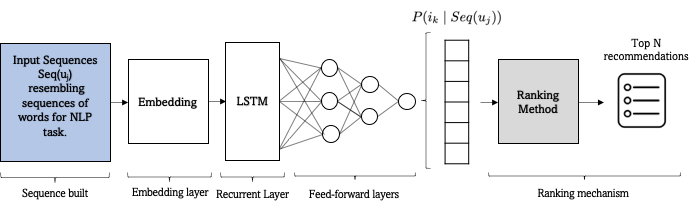}
\caption{Proposed recurrent model to estimate probability of interactions of future tokens. The input transactional sequences $Seq(u_{j})$ are processed through an embedding layer and two Long-Short Term Memory cell (LSTM) followed by a Multilayer Perceptron with sigmoid  activation of size $\mid I' \mid$ to obtain the probabilities of user-item interaction $P_{u_{j}}(i_{k})$.}

\label{fig:RNN_model}
\end{figure*}

To estimate the probability $P_{u_{j}}(i_{k})$ we use the recurrent neural network shown in Figure \ref{fig:RNN_model}, which consists of an embedding layer followed by two Long-short Term Memory  (LSTM) units and a five-layer feed-forward network with sigmoid activation and output size $\mid I' \mid$ corresponding to all potentially recommendable items.

\subsection*{Data Pre-processing}  \label{data_preprocessing}

Inspired by the approach that NLP techniques take to compress and process sequential data, we aim to process user-items interactions as a sequential task. To achieve this, we take the original transactional data shown in Table \ref{table:Original_transactional_data}, which is typically used to build the user-item matrix presented in the last section, and build a sequence of item interactions $Seq(u_{j}) = [i_{u_{j},1} \, i_{u_{j},2}  \,  ... \, i_{u_{j},u_{n_{j}}}]$ for every user $u_{j} \in U$ as shown in table \ref{table:Sequential_transactions}, where the elements of this list are the space-separated and timestamp-ordered item ids that the user $u_{j}$ has interacted with. Like this, it is straightforward to process the sequence $ Seq(u_{j})$ as a textual sequence where each item id would simply resemble a single word in a dictionary for an NLP task. 

\begin{table}[!htb]
      \caption{Original Transactional data}
      \centering
        \begin{tabular}{ccc}
        \toprule
        User ID & Item ID & timestamp \\ \midrule
        $u_{1}$ & $i_{1}$ & $t_{1,0}$ \\ 
        $u_{1}$ & $i_{2}$ & $t_{1,1}$ \\ 
        $u_{2}$ & $i_{1}$ & $t_{2,0}$ \\ 
        ... & ... & ... \\
        $u_{j}$ & $i_{d}$ & $t_{j,d}$ \\
        $u_{j}$ & $i_{d+1}$ & $t_{j,d+1}$ \\
        ... & ... & ... \\ \bottomrule
        \end{tabular}
        \label{table:Original_transactional_data}
\end{table}

\begin{table}[!h]
      \centering
        \caption{Customer sequential-transactions}
            \begin{tabular}{cc}
            \toprule
            Customer ID & $Seq(u_{j})$ \\ \midrule
            $u_{1}$ & [$i_{u_{1},1} \, i_{u_{1},2} \, ... \, i_{u_{1},n_{1}}$] \\ 
            $u_{2}$ & [$i_{u_{2},1} \, i_{u_{2},2} \, ... \, i_{u_{2},n_{2}}$] \\ 
            ... & ... \\ 
            $u_{j}$ & [$i_{u_{j},1} \, i_{u_{j},2} \, ... \, i_{u_{j},n_{k}}$] \\
            ... & ... \\ \bottomrule
            \end{tabular}
            \label{table:Sequential_transactions}
\end{table}

Then, the target vector for the model for a user $u_{j}$ is built as $Y_{u_{j}} = [y_{u_{j},i}, y_{u_{j},2}, … , y_{u_{j}, \mid I' \mid}]$, where $y_{u_{j}, k} = 1$ if the user $u_{j}$ interacted with the item $i_{k}$ over the performance period, and $0$ otherwise.

\subsubsection*{Tokenisation and Embeddings} \label{section_3_1_1}
\subsubsection*{Tokenisation} \label{Tokenization}
Tokenisation is a simple but powerful technique in NLP to transform a sentence of text and split it into the different elements or ‘tokens’ that compose it. For example, given the textual sentence \emph{‘recommendations for different users and items’}, the tokenisation returns a sequence of six tokens \emph{[‘recommendations’ ‘for’ ‘different’ ‘users’ ‘and’ ‘items’]}, with each single word as its own token. \emph{Dictionary-based tokenisation} is perhaps the most common type of tokenisation used in the AI industry, this method uses a pre-defined dictionary of mapped words into tokens, typically learnt from a large set of textual sequences, which allows to tokenise every new given sentence to be processed. However, there are different tokenisation methods such as \emph{rule-based tokenisation}, \emph{regular expression tokenisation}, or \emph{sub-word tokenisation} in the NLP literature that are not covered in the scope of this work. 

As in this work the tokens to be processed do not correspond to real words in a pre-trained dictionary but to item identifiers, it is necessary to learn a new dictionary from the data during model training and use it to compress the sequences of purchases at prediction phase.

\subsubsection*{Word2Vec Embeddings} \label{Embeddings}
Word embeddings are powerful methods used in natural language processing and information retrieval to overcome the high-dimensionality problem of dealing with large corpus of text by obtaining representations of words contained in documents \cite{Bengio_2003}. Word2Vec embeddings were introduced by \cite{ Mikolov_2013} in 2013 as an approach to learn high-quality dense vector representations of words for dictionaries with potentially billions of tokens, while trying to keep similarity of words in terms of their semantics and position within sentences. 

In particular, the \emph{Continuous Bag-of-Words} (CBOW) \cite{ Mikolov_2013} is an unsupervised neural network with a single fully connected hidden layer (a.k.a shared-projection matrix) for all words to represent each token in the network and predict the current word based on its context or surrounding words. Despite its simplicity, CBOWs are popularly used in textual applications to find scalable word representations as an alternative to the highly sparse bag-of-words representation. Although Word2Vec embeddings are considered unsupervised methods, these are learned as part of a supervised framework such as classification or next token prediction, as in this work, by simply connecting the embedding layer within the overall architecture and training via backpropagation as shown in figure \ref{fig:RNN_model}. 


\subsubsection*{N-gram Models} \label{n_grams}
N-gram models were proposed in the field of NLP to efficiently predict the next token in a sentence given the last \emph{n-1} tokens, these techniques have been widely used in applications like spell-checking, text generation, and DNA sequencing \cite{Bickel_2005, Wang_2012} to capture the representation of token distributions throughout sequences assuming contextual correlations, i.e., closest tokens have strong correlation within each other than far tokens. 

In its simplest form, token predictions are computed via maximum likelihood estimation by estimating $P(w_{n}) = \prod_{k=1}^n P(w_{k} \mid w_{1:k-1})$ where $w_{k}$ represents the \emph{k-th} token in a given sequence \cite{Bickel_2005}. However, several approaches add the use of \emph{clustering} of tokens into different groups and \emph{smoothing} of the probability of each token with \emph{mixture-models}, or the use of encoding-decoding and attention mechanisms aiming to increase prediction performance \cite{Qi_2020}. 

\subsubsection*{Loss function} \label{section_loss_function}
For model training we use binary cross-entropy as loss function and stochastic gradient descent with backpropagation for optimisation of the model architecture shown in Figure \ref{fig:RNN_model}, therefore the optimisation problem can be written as minimising the loss function defined as 

\[
Loss = - \frac{1}{\mid U \mid} \sum_{j=1}^{\mid U \mid} \sum_{i=1}^{\mid I' \mid} y_{i,j} \cdot log(P_{u_{j}}( i_{k} ))  
\] 

where $y_{i,j}$ is the binary target variable introduced in section the data pre-processing section which corresponds to the indicator function of user-item interaction over performance period, and $P_{u_{j}}( i_{k} )$ is the estimation of the probability of the neural network.

\subsubsection*{Ranking Mechanism} \label{ranking_method}
As the final goal of a recommender system is to output a list of potentially relevant items for users, a wide range of techniques incorporate ranking methods and ad-hoc ranking losses during training to stream model training and testing and mitigate serving biases \cite{Covington_2016, Zhang_2019, Shani_2005}. The ranking process also allows incorporation of additional business-related information about the items without affecting the estimation of probabilities, for example, in a real-world scenario we might be interested in recommending an item which has slightly lower probability of interaction with users, but that it is more profitable for the company that is making recommendations. 

In this work we keep the ranking process separated from the overall model architecture at training, as it allows higher flexibility to react to business rules and item offers that can be promoted by companies. The process to obtain the final item list is straightforward by recommending the items which obtain the highest uplift $R(u_{j},i_{k})$ in terms of user-item interaction probability against the baseline probability of interactions for an item $i_{k}$.

\[
R(u_{j},i_{k}) = \frac{P_{u_{j}}(i_{k})}{P(i_{k})}
\]
\medskip

where $ P(i_{k})$ represents the probability that a random user interacts with item $i_{k}$. $R(u_{j},i_{k})$ can be interpreted as how many times a user $u_{j}$ is more likely to interact with item $i_{k}$ against a random user. 

\subsubsection*{Training process} \label{Training}
The training process of the neural network shown in Figure \ref{fig:RNN_model} is carried by minimising the loss function introduced previously via backpropagation. To create the input sequences, we decided to split transactions into two disjoint sets to obtain the observation and performance data with respect to an arbitrary analysis date which can be set according to the business needs and frequency of interactions.  Then input sequences $Seq(u_{j})$ can be created for all users by using only transactions contained in the observation period (prior to the analysis date), whereas targets are built out of transactions in performance (after the analysis date), as shown in in the data pre-processing section. Once sequences are obtained for all users, we split further this data into 80\% of the sequences for model training and 20\% for validation of results. Pseudo-code shown in Algorithm \ref{alg:Training_pseudo_code} details the full procedure for data pre-processing and model training. 

\begin{algorithm}
\caption{Sequence-aware recommender system pseudo-code.} \label{alg:Training_pseudo_code}
\begin{algorithmic}
\State \textbf{Input:} Customer transactional data presented in table \ref{table:Original_transactional_data} including \emph{'Customer id'},  \emph{'Purchase date'}, \emph{'Item id'}. \\
\State \textbf{Output:} Item list recommendations for users. 
\\

\State \textbf{Data Split:}
\State - Split training w.r.t the set analysis date to get observation and performance transactional data.
\State - Split total data into two disjoint sets of customers for training (80\%) and validation (20\%).

\\

\State \textbf{Data Pre-processing:}
\State - With the observation transactional data, for each user in the dataset create $Seq(u_{j})$ with the time-ordered product ids of previous items purchased by user $u_{j}$  as stated in the data pre-processing section.
\State - With performance data, for each user create the target vector 
\[Y_{u_{j}} = [y_{u_{j},i_{1}}, y_{u_{j},i_{2}}, … , y_{u_{j}, i_{\mid I \mid}}],\]
with user-item interactions in the performance period as stated in data pre-processing section.
\\ 

\State \textbf{Model Training:}
\State - Train the sequence-aware model via backpropagation with the architecture presented in figure \ref{fig:RNN_model}, the data obtained from the pre-processing step, and the binary cross-entropy outlined in the loss function section as loss function.
\\

\State \textbf{Model Prediction:}
\State - For each user, obtain the probability of interaction with each potentially recommendable item $\hat{P}_{u} = [\hat{P}_{u_{j}}(i_{1}),..., \hat{P}_{u_{j}}(i_{\mid I \mid})]$. 
\\

\State \textbf{Ranking Process:} 
\State - For each user $u_{j}$ and item $i_{k} \in \mid I' \mid$, obtain the estimated uplift $R(u_{j},i_{k})$ in probability interaction as stated in the ranking mechanism section
\[
R(u_{j},i_{k}) = \left[ \frac{\hat{P}_{u_{j}}(i_{1})}{P(i_{1})},..., \frac{\hat{P}_{u_{j}}(i_{\mid I' \mid})}{P(i_{\mid I' \mid})} \right]
\]
\State - For each user, recommend the top K items for which $R(u_{j},i_{k})$ is the largest.
\end{algorithmic}

\end{algorithm}

\subsection*{Performance Metrics} \label{section3.3}
Evaluating recommender systems introduces an extra level of complexity compared to evaluating traditional classification or regression techniques for a number of reasons: The potential absence of knowledge about the real user preference over items in implicit feedback settings may induce algorithmic bias in model evaluation, as it is not possible to compare model predictions against the explicit user-item rating via error metrics like MSE used in matrix-completion techniques. Different techniques used to perform recommendations may also perform significantly differently depending on some characteristics of the dataset like number of users and items, data sparsity, rating scale, among others. This may lead to having to depend on scenario-specific metrics and manual analysis for evaluating recommendations. Finally, popular evaluation metrics used for recommender systems completely ignore the fact that most users have not had the opportunity to interact with products due to unfamiliarity to them, and not due to a lack of preference. Furthermore, these metrics are designed only for off-line evaluation from previous user-item interactions and might not be scalable to production settings where efficiency must be tested immediately, and designing on-line evaluation experiments to assess recommendation performance might require much more effort and usually these are more expensive. 

\cite{ Fouss_2009} propose performance evaluation of recommender systems by taking into account different characteristics like coverage, to measure the percentage of the dataset for which the recommender is able to provide recommendations, computing time to measure how quickly the system can make recommendations for large set of users, and robustness to assess how good the model is in presence of added noise in the data. 

For the purpose of this work, some of these metrics might or might not apply in our case, as we are trying to estimate the probability of user interaction with items, rather than item relevance or how much the user will like each item. The following sections briefly outline the performance metrics used to evaluate model performance and user recommendations for our approach from a technical perspective.

\subsubsection*{Normalised Discounted Cumulative Gain}
Introduced by \cite{ rvelin_2002}  the normalised discounted cumulative gain (NDCG) is a popular ranking quality metric widely used in information retrieval and recommender systems to evaluate list of items with length $p$, it takes into account the degree of relevance of items via an information gain function (Discounted Cumulative Gain) defined as 

\[
DCGp = \sum_{k=1}^{p} \frac{relevance_{i_{k}}}{log_{2}(k+1)},
\]

as well as the ranking of relevant items in the list via a discount function with respect to the best ranking possible (Ideal Discounted Cumulative Gain).
\[
IDCGp = \sum_{k=1}^{REL_{p}} \frac{relevance_{i_{k}}}{log_{2}(k+1)},
\]
where $REL_{p}$ represents the list of all possible relevant items up to position $p$ and $relevance_{i_{k}}=1$ if the item is considered relevant for the user, and 0 otherwise. Thus, the normalised discounted cumulative gain is defined as 

\[
NDCG_{p} = \frac{ DCGp }{ IDCGp }, 
\]
which ranges from 0 to 1, and where higher values of NDCG are associated with a better ranking of items in the recommendation list.

\subsubsection*{Mean Average Precision}
The Mean Average Precision (MAP) is a metric originated in the field of information retrieval, it provides insight about how relevant a list of items is with respect to all possible user queries. In recommender systems evaluation, this metric evaluates how good recommendation lists of $K$ items are by obtaining the mean of the Average Precision (AP@K) for each each list defined as 
\[
AP@K = \frac{1}{rel_{K}} \sum_{k=1}^{K} \frac{\text{\# of relevant items at k}}{k},
\]
where $rel_{K}$ is the number of total relevant items in the top $K$ results, and can be interpreted as the proportion of relevant items for the user in the recommendation list.

\subsubsection*{Sales and Revenue} \label{section_sales_revenue}
As mentioned in the first section, one of the main goals of recommending the right items to user is aiming to increase overall sales and revenue of the businesses \cite{Gunawardana_2009}, this can be either by cross-selling relevant items during the purchase order, by renewing a subscription to a service due to the highly personalised offering, or by displaying advertising with the right offers to customers to encourage them to take an action, such as purchase an item that they were not thinking in buying or by staying shopping longer. 

In this work, we aim to predict the probability of future interaction between users and items to offer the most likely items to purchase first, regardless of the users already having interacted with the item in the past. Therefore, encouraging customers to purchase the right set of items that businesses want to promote. To assess the uplift on revenue in our methodology against other recommender systems, of which we cannot provide any technical detail due to intellectual property constrains, we designed a live A/B test experiment with a large retailer in the UK to assess the impact in revenue of making recommendations with our method against other recommender systems approaches. Results of this experiment are presented in the next section.

\section*{Experiments and Results} \label{section4}
Profusion is a data consultancy company based in London, UK, that provides data analytics services to retailers and financial companies across the UK. To increase revenue and improve customer engagement, Profusion uses a wide range of techniques to perform item recommendations. In order to assess the effectiveness of the system, we conducted several experiments on two real-world retail datasets provided by Profusion clients. For confidentiality purposes, names of these two retail companies will not be displayed and we reefer to them as \emph{company 1} and \emph{company 2}. 

\subsection*{Dataset 1: \emph{Company 1 - Retail}}
The first dataset contains transactions made by customers of \emph{company 1}, a retailer in the UK specialised in selling alcoholic beverages. The total volume of transactions in the sampled data consists of nearly 80 millions of purchases made by over 3 million customers between March 2016 and February 2022. Although there are 10,000 different items available for customers over the whole observation period, on average, customers only interact with 24 of these items over their whole customer life cycle, and rarely provide explicit feedback about the purchased items, which induces several issues related to training recommender systems due to the highly sparse data and the lack of any explicit feedback.  

For our off-line experiments we used 80\% of the total transactional data prior to a defined analysis date (September 2021), allowing to have 6  months of customer transactions for performance, and the 100\% of the data after the analysis date for evaluation purposes. Results of the tests for \emph{company 1} are presented in Table \ref{table:Performance_results}.
 
\subsection*{Dataset 2: \emph{Company 2 - Retail}}
The second dataset contains information of transactions in \emph{company 2}, a large building equipment retailer in the UK with nearly 80 million of purchases made by 1.2 million customers between June 2020 and March 2022. The dataset contains 73,000 potentially recommendable items, although on average each user interacts with only 51 different items over their life cycle. 

Similarly to the first dataset, we consider only 80\% of customer transactional data prior to a defined analysis date (September 2021) to allow a 6 months performance window, and the rest of transactions for model testing. Model performance for this dataset is presented in Table \ref{table:Performance_results}.

\subsection*{Dataset 3: \emph{Open dataset: Movielens 25M}}
The MovieLens dataset \cite{movielens} is widely used in research and industry to benchmark recommender systems performance. The version of the data used in this work, MovieLens 25M provided by \emph{GroupLens Research}, contains 25 million movie ratings for 62,000 movies and 162,000 users during January 1995 and November 2019. Each of the users in this data have rated at least 20 movies, which mitigates the cold-start problem of not having enough information. As mentioned previosuly, the data pre-processing needed does not make use of ratings, thus we only consider the interaction between users and items in the data.  Model performance for the MovieLens dataset is presented in Table \ref{table:Performance_results}.

\subsection*{Results} \label{sec_results}

In this section we empirically demonstrate the effectiveness of our approach by comparing the ranking metrics presented previously: MAP@1, MAP@10, and NDCG for the three mentioned datasets in the retail industry against collaborative filtering and matrix factorisation, Table \ref{table:Performance_results} include the details of this performance assessment. Additionally to the off-line evaluation, we evaluate the impact in overall revenue by conducting a live A/B experiment for \emph{Company 1} at comparing sales made with our method against another recommender system based on matrix factorisation used by \emph{Company 1}, further detail of this test is presented in the next section in table \ref{Tab:Performance_results_A_B}.

It is worth mentioning that metrics obtained in this section, MAP and NDCG are computed only from previous purchases made by customers over a period of six months, which do not take into account the fact that customers might not be aware of the existence of all the possible recommendable items in the product catalogue and have only knowledge of popular items, this might induce a bias of equal opportunity in assessing performance of user-item interaction for unpopular items in recommender systems, at these items being inherently less likely to be purchased by users even when in cases where the items are highly relevant. To evaluate our method against CF and MF techniques, while overcoming this issue, we conducted off-line recommendations by training all recommender systems with a sampled dataset which do not considers purchases of the top 10\% more popular items in each dataset. Performance results of MAP@1, MAP@10, and NDCG for this experiment are presented in Table \ref{table:Performance_results_sampled_data}. 

\begin{table*}[!htbp]
\caption{Recommendation performance obtained from off-line experiments with 20\% of unobserved customers at training in the validation dataset and allowing 6 months of user-item interactions as the performance period. }
\centering
\resizebox{0.95 \textwidth}{!}{
    \centering
    \begin{tabular}{lccccccccc}
     & \multicolumn{3}{c}{\textbf{\emph{Company 1}}} & \multicolumn{3}{c}{\textbf{\emph{Company 2}}} & \multicolumn{3}{c}{\textbf{\emph{MovieLens}}} \\ 
        \midrule
         & \textbf{MAP@1} & \textbf{MAP@10} & \textbf{NDCG} & \textbf{MAP@1} & \textbf{MAP@10} & \textbf{NDCG} & \textbf{MAP@1} & \textbf{MAP@10} & \textbf{NDCG} \\ 
        \midrule
        Sequence-aware & \textbf{0.0119} & 0.0233 & \textbf{0.0314} & \textbf{0.0354} & 0.1003 & \textbf{0.1559} & \textbf{0.0085} & \textbf{0.0124} & \textbf{0.0160} \\ 
        Collaborative Filtering & 0.0111 & \textbf{0.0265} & 0.0215 & 0.0139 & 0.1016 & 0.0607 & 0.0016 & 0.0046 & 0.0067 \\ 
        Matrix Factorisation & 0.0014 & 0.0144 & 0.0265 & 0.0146 & \textbf{0.1143} & 0.0643 & 0.0021 & 0.0046 & 0.0063\\ 
        \bottomrule
    \end{tabular}}
    \label{table:Performance_results}
\end{table*}

\begin{table*}[!htbp]
\caption{Recommendation performance from a subset training dataset obtained by removing the 10\% most popular items at training and evaluating performance for 20\% unobserved customer in the validation dataset with a 6 months of transactions in the observation window.}
\centering
\resizebox{0.95 \textwidth}{!}{
    \centering
    \begin{tabular}{lccccccccc}
     & \multicolumn{3}{c}{\textbf{\emph{Company 1}}} & \multicolumn{3}{c}{\textbf{\emph{Company 2}}} & \multicolumn{3}{c}{\textbf{\emph{MovieLens}}} \\ 
        \midrule
         & \textbf{MAP@1} & \textbf{MAP@10} & \textbf{NDCG} & \textbf{MAP@1} & \textbf{MAP@10} & \textbf{NDCG} & \textbf{MAP@1} & \textbf{MAP@10} & \textbf{NDCG} \\ 
        \midrule
        Sequence-aware & \textbf{0.0051} & 0.0078 & \textbf{0.0095} & \textbf{0.0050} & 0.0092 & 0.0141 & 0.0010 & 0.0081 & 0.0177 \\ 
        Collaborative Filtering & 0.0015 & \textbf{0.0091} & 0.0058 & 0.0000 & \textbf{0.0186} & 0.0013 & \textbf{0.0210} & \textbf{0.0403} & \textbf{0.0535} \\ 
        Matrix Factorisation & 0.0026 & 0.0055 & 0.0069 & 0.0000 & 0.0053 & \textbf{0.0162} & 0.0000 & 0.0130 & 0.0090\\ 
        \bottomrule
    \end{tabular}}
    \label{table:Performance_results_sampled_data}
\end{table*}

\subsubsection*{Live A/B test results.}\label{live_results}
\begin{figure}[!htbp]
\centering{}
\includegraphics[scale=0.45]{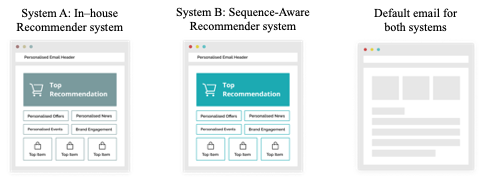}
\caption{Email template sent to customers in live A/B test. 50\% of customers were selected for recommendation of system A using Matrix factorisation and the rest 50\% with sequence-aware recommendations. Only customers without transaction history were selected to be included as part of the default template with popular items.}
\label{fig:email_banner}
\end{figure}

As mentioned, evaluating of recommender systems is usually carried with off-line metrics such as Mean Average Error (MAE) when information about user ratings is known, and ranking metrics like MAP@K and NDCG when there is no information available about real user preferences. Additionally to the off-line evaluation carried for MAP@K and NDCG, we conducted a live A/B test item for item recommendations included in an email marketing campaign for \emph{company 1}. In this test, we tested our method against an in-house recommender system for which we cannot provide further details other than that overall performance compares to a matrix factorisation model. This test only considered the top 1 item recommendation from a universe of 8 recommendable items that the company wanted to promote in an email marketing campaign for over 500,000 customers. Six of these items are considered priority for \emph{company 1} and should be ranked first if possible, the other two items are considered popular items and should be sent as a default in case there is no better recommendation for customers. The top 1 recommendation for both systems, as well as the default items were presented as the main banner in the email sent to customers, as illustrated in Figure \ref{fig:email_banner}, and the rest of the email components such as the header, images, and any other items promoted in the email remained the same for both systems.

\begin{figure*}[!htbp]
\centering{}
\includegraphics[scale=0.70]{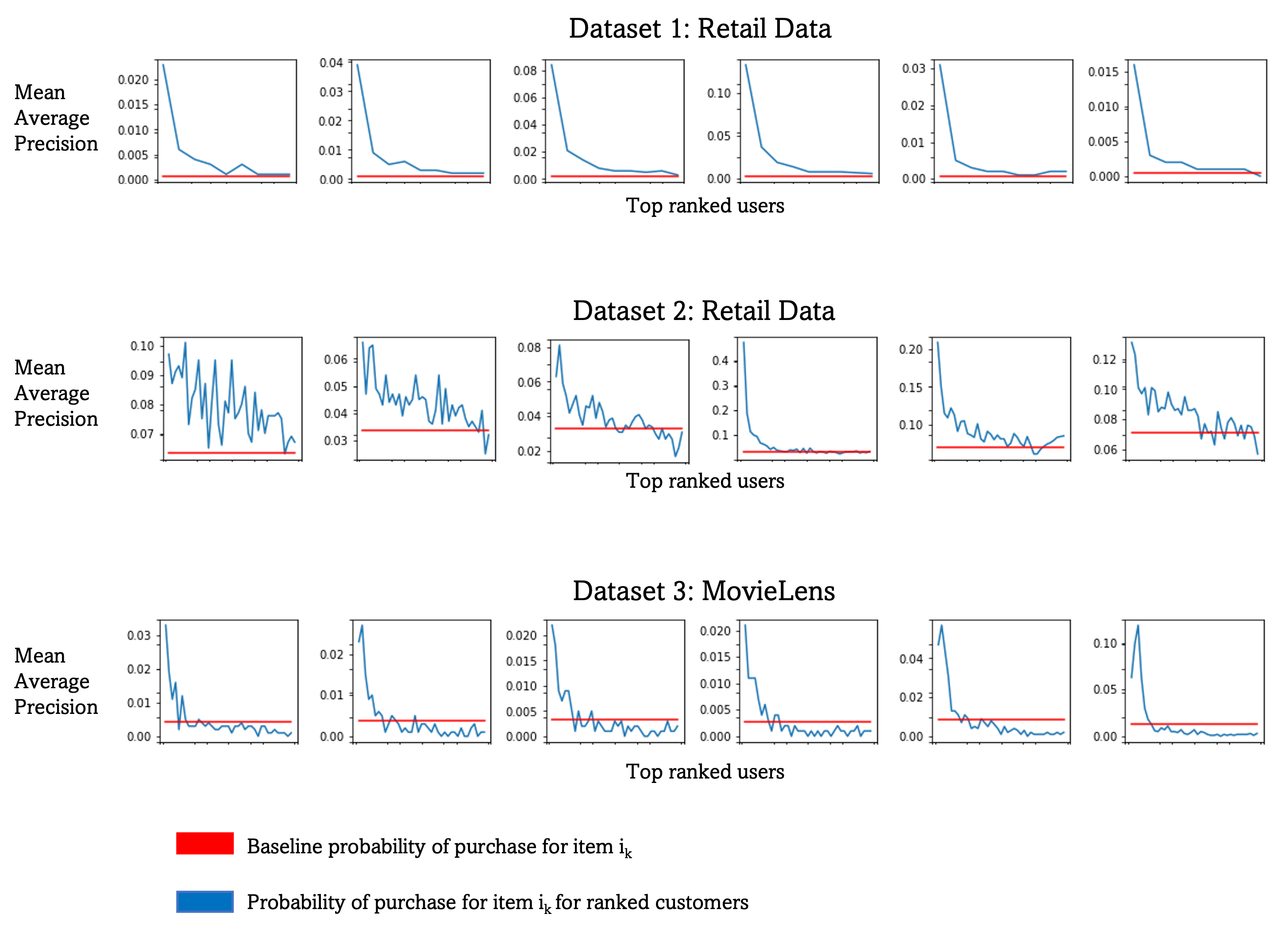}
\caption{Mean Average Precision for Sequence-aware recommendations for 6 products in each dataset selected at random.}
\label{fig:rank 8 items}
\end{figure*}

Although we cannot compare these systems at ranking level due to confidentiality constraints, we can provide the average ranking probability of purchase for each in the selection of  eight items used for the live test. As shown in Figure \ref{fig:rank 8 items}, our recommender system could identify the best item to be recommended to customers and rank customers with 20-40 times higher precision at the top of the ranking list. Besides the overall sales results showed that the average customer revenue for users targeted with our system increased by 51\% in comparison to users revenue targeted by the company system. Further detail of the A/B test revenue results is presented in Table \ref{Tab:Performance_results_A_B}. 

\begin{table*}[!htbp]
\centering
\caption{Revenue results of A/B test carried for \emph{Company 1} using our  sequence-aware method to make item recommendations against an in-house recommender system based on matrix factorisation. Due to confidentiality constraints with the company providing data and settings for the experiments, sales amounts multiplied by a random number to be presented in this report and preserve confidentiality.}
\resizebox{0.99 \textwidth}{!}{
\begin{tabular}{lllllllll}
 & \multicolumn{4}{c}{System A: In-house recommender system} & \multicolumn{4}{c}{System B: Sequence-Aware recommender system} \\
  \midrule
Item & Volume of Users & Response & Total Revenue & Revenue p/c & Volume of Users & Response & Total Revenue & Revenue p/c \\
 \midrule
item 1 & 7.0\% & 1.75\% & £13,622 & £37.73 & 4.4\% & 0.99\% & £9,030 & £69.44 \\
item 2 & 8.8\% & 1.17\% & £13,720 & £45.15 & 8.9\% & 0.95\% & £20,440 & £81.76 \\
item 3 & 3.7\% & 0.88\% & £2,149 & £22.12 & 16.4\% & 0.88\% & £24,472 & £57.05 \\
item 4 & 11.5\%  & 1.81\% & £35,105 & £56.70 & 8.1\% & 1.02\% & £21,042 & £85.54 \\
item 5 & 8.2\% & 1.49\% & £17,003 & £46.97 & 9.2\% & 0.88\% & £20,517 & £85.47 \\
item 6 & 10.6\% & 0.88\% & £14,294 & £51.80 & 6.1\% & 0.99\% &12,719 & £70.28 \\
item 7 & 9.8\% & 1.27\% & £16,079 & £43.54 & 38.9\% & 0.48\% & £25,578 & £46.62 \\
item 8 & 40.5\% & 0.60\% & £15,981 & £22.33 & 7.9\% & 0.56\% & £672 & £5.11 \\
 \midrule
Total & 100\% & 1.05\% & £127,953 & £41.23 & 100\% & 0.73\% & £134,470 & £62.37 \\
 \bottomrule
\end{tabular}}
\label{Tab:Performance_results_A_B}
\end{table*}

\section*{Conclusion} \label{section5}

This work presents an innovative approach to make item recommendations for individual customers while considering the order than previous purchases were made, all by using recurrent neural networks and data pre-processing methods traditionally used in the field of natural language processing to make predictions of words. Although sequence-aware recommender systems have been widely explored and used in fields where customer preferences change drastically in short periods of time, such as mobile app usage, and video games applications such as the ones mentioned in second section, to the best of our knowledge there are no applications of sequence-aware item recommendations for the retail industry as the ones presented in the datasets used in the fourth section. 

The main motivation of this work is performing item recommendations in settings such as retail where matrix-completion like collaborative filtering and matrix factorisation approaches presented in the second section do not perform well for to several different reasons, while recommendation performance is at least still similar to the mentioned and widely used methods, specially for the top of the recommendation list. For instance, user preferences might change rapidly over time, or users may interact repeatedly with specific items without providing explicit ratings. 

The proposed sequence-aware recommender system in this work overcome these issues and maintain predictive performance comparable to other recommendation techniques by inherently assuming that items may be present in the purchasing sequence multiple times, and all items, including the ones previously bought by users, may be contained in the set of recommendable items. This change allows us to achieve two main goals: firstly, improve customer engagement by identifying the best products for users, regardless of whether these have been already purchased by users. Secondly, improve customer journey by identifying which items are most likely to be purchased in sequential order, and provide tailored recommendations with special offers. 

Additionally, the cost of maintaining our recommender system in production environments is lower than maintaining matrix completion methods, at these need to be constantly updated to account for new user-item interactions, whereas our method only considers this change in the input data fed into the network, which is extremely convenient for businesses and machine learning engineers. Furthermore, our method generates item recommendations considerably faster than matrix completion techniques, at not needing to estimate preferences for all users and all items simultaneously, but one user at the time instead. 

However, although our method has achieved great performance in compared to the baselines models outlined in in the fourth section, it still has its limitations. As the final probability for user-item interactions is estimated from the soft-maxed output layer of the recurrent neural network, our methodology is prone to suffer selection bias induced from highly popular items. This effect occurs at observing an extremely imbalanced number of interactions for a specific item during training, which might not persist at prediction phase, and therefore affecting the item ranking overall. Additionally, our methodology suffers severely from cold start problems at introducing new items for ranking, as training targets introduced in section the data pre-processing are built from a specific period for each item. Finally, as shown in table \ref{table:Performance_results_sampled_data}, our method might not be the best performing where the size of the recommendation list for each customer is large, but rather in recommendation scenarios where the goal is finding the top 1 best item to recommend based in the most recent data. 

Our methodology presented in this work could be easily extended in different ways to improve item recommendation performance, as the main focus of our approach is just encoding and capturing information of previous purchases for each customer, we are so far ignoring information from users and items characteristics, as well as potential hand-crafted features that can be easily added to the input of the neural network or to any other model that could be used to perform token prediction.


%




\backmatter
\newpage
\section*{Declarations}

\bmhead{Funding}
This work was supported by the Knowledge Transfer Partnership program through Innovate UK, University of Essex (Registration Number: Z699129X) and Profusion Media LTD (a company registered in England, number 6947442).

\bmhead{Competing interests}
The authors have no relevant financial or non-financial interests to disclose.

\bmhead{Code availability}
The code for this work is intellectual property of Profusion Media LTD. and not publicly available. 

\bmhead{Authors' contributions}
J.E implemented the algorithms, conducted the data analysis and wrote the initial draft, and B.L. verified the results and supervised this work. All authors revised the paper. All authors read and agreed to the published version of the manuscript.




\bigskip
\newpage

\bibliographystyle{plain}
\bibliography{biblio}

\end{document}